\documentstyle[twoside]{article}

\def\nn{\nonumber\\}
\def\nq{\hspace{-1em}}
\def\nqq{\hspace{-2em}}
\def\nhq{\hspace{-0.5em}}

\def\cm{\hspace{1cm}}

\def\al{&\nhq}
\def\noi{\noindent}

\def\beq{\begin{equation}}
\def\eeq{\end{equation}}
\def\bear{\begin{eqnarray}}
\def\ear{\end{eqnarray}}
\def\eql{\al =\al}
\def\lal{&&\nqq}     

\newcommand{\Acknow}[1]{\bigskip{\bf Acknowledgement}\\[1ex] \noindent #1}

\newcommand{\bls}[1]{\renewcommand{\baselinestretch}{#1}}

\def\const{{\rm const}}

\def\e{{\rm e}}
\def\to{\rightarrow}
\def\gg{\overline{g}}
\def\RR{\overline{R}}
\def\ds{d\overline{s}^2}
\def\half{{\textstyle\frac 12}}
\def\SAS{static, axially symmetric\ }
\def\kR{\Bigl(1-\frac{2k}{R}\Bigr)}

\newcommand{\vars}[1]{\left\{\begin{array}{ll}#1\end{array}\right.}

\bls{1.10}
\begin{document}
\thispagestyle{empty}
{\large
\begin{center}
               RUSSIAN GRAVITATIONAL SOCIETY\\
     CENTER FOR GRAVITATION AND FUNDAMENTAL METROLOGY, VNIIMS
\end{center}
\vskip 4ex
\begin{flushright}                 RGS-VNIIMS-003/95\\
                                   gr-qc/9505021
\end{flushright}
\vskip 15mm

\begin{center}
{\Large\bf VACUUM STATIC, AXIALLY SYMMETRIC FIELDS\\[5pt]
     IN D-DIMENSIONAL GRAVITY}

\vskip2.5ex
     {\bf K.A. Bronnikov and V.N. Melnikov}\\
\vskip 5mm
     {\it Center for Gravitation and Fundamental Metrology, VNIIMS\\
     3--1 M. Ulyanovoy Str., Moscow 117313, Russia}\\
     e-mail: bron@cvsi.rc.ac.ru {\it or\ } mel@cvsi.rc.ac.ru\\
\vskip 10mm
\end{center}

\noi{\bf ABSTRACT}

\bigskip\noi
Vacuum \SAS space-times in $D$-dimensional general relativity with a
Ricci-flat internal space are discussed. It is shown, in particular, that
some of the monopole-type solutions are free of curvature singularities and
their source can be a disk membrane bounded by a ring with a string or
branching type singularity. Another possibility is a wormhole configuration
where a particle can penetrate to another spatial infinity by passing through
a ring with a string or branching type singularity. The results apply, in
particular, to vacuum and scalar-vacuum configurations in conventional
general relativity.

\vfill
\centerline{Moscow 1995}
}
\pagebreak
\twocolumn

\section{Introduction}
\markboth{1. Introduction}{1. Introduction}
     Spherically symmetric static solutions of multidimensional gravity
     have been considered by many authors with a goal to study possible
     observational windows to extra dimensions [1--3]. Among such windows
     one can name possible variations of fundamental physical constants
     [4], deviations from Newton's and Coulomb's laws and modified
     properties of black holes and gravitational radiation as compared
     with the conventional theory.

     Another class of multidimensional models, to be discussed in this
     paper, is the class of axially symmetric models, including
     spherically symmetric ones as a special case.

     Although \SAS (SAS) configurations are a less popular object of
     gravitational studies than stationary ones (used for describing
     fields due to rotating bodies), their properties are of much interest
     as well. In many papers such solutions are sought and studied, see,
     for instance, \cite{Lya,Rad77,Synge} and references therein.
     We will study monopole SAS vacuum configurations in multidimensional
     gravity and find some features of interest, in particular, membrane
     and string type sources of fields possessing no curvature
     singularities.

        We consider $D$-dimensional general relativity and start from the
action
\beq
        S = \int d^D x \sqrt{^D g}(^D R + L_m)              \label{Ac1}
\eeq
        where $L_m$ is a matter Lagrangian, in a space with the metric
\beq
        ds_D^2 = g_{\mu\nu}dx^{\mu}dx^{\nu}
                  + \e^{2\beta_1}ds_1^2                        \label{DsD}
\eeq
        where Greek indices range from 0 to 3 and $\beta_1(x^{\mu})$ is a scale
        factor of an internal $N$-dimensional space with a Ricci-flat
        $ds_1^2$ independent of $x^{\mu}$.

        In a 4-dimensional formulation
\beq
        S = \int d^4 x \sqrt{^4 g}\e^{\sigma}
     \Bigl[R - \Bigl(\frac{1}{N}-1 \Bigr)\sigma^{\alpha}\sigma_{\alpha}
                           +L_m\Bigr]                \label{Ac2}    
\eeq
     where $\sigma = N\beta_1$ and $R$ is
     the 4-curvature corresponding to $g_{\mu\nu}$.

        There are other 4-dimensional formulations of the theory, connected
        with (\ref{Ac2}) by conformal mappings ({\it conformal gauges}).
        The gauge (\ref{Ac2}) corresponds to the
        original theory. The so-called {\it Einstein gauge}, obtained
        from (\ref{Ac2}) by the conformal mapping
\beq
        \gg_{\mu\nu} = \e^{\sigma} g_{\mu\nu},                  \label{GE}
\eeq
        is more convenient for solving the field equations
        since the curvature enters into the Lagrangian with a constant
        factor:
\bear
        S \eql \int d^4 x \sqrt{^4 \gg}\e^{\sigma}
        \bigl[\RR + \alpha_0 \gg^{\alpha\beta}\sigma_{\alpha}\sigma_{\beta}
          +\e^{-\sigma}L_m \bigr],                  \nn
          \lal\quad  \alpha_0 = 1/2 + 1/N.               \label{Ac3}
\ear
     where $\RR$ is the scalar curvature corresponding to $\gg_{\mu\nu}$.
     Another important gauge, the so-called {\it atomic\ } one, in which
     a test particle moves along geodesics, is defined by
\beq
        g^*_{\mu\nu} = \e^{\sigma/2} g_{\mu\nu}          \label{GA}   
\eeq
     and is most suitable for interpretaion of measurements, e.g.,
     in the Solar system. However, for
     studies of singularities and topology the original metric
     $g_{\mu\nu}$ must be used. For more detailed discussion of the
     notion of systems of measurement, closely connected with that of
     conformal gauges, see Ref.\cite{StM} and, as
     applied to multidimensional theory, Refs.\cite{Br1,Br2,MRF}.

     In what follows we use the Einstein gauge
     to find the metric (\ref{GE}) for vacuum SAS configurations.
     So we start from the equations due to (\ref{Ac3}) with $L_m=0$:
\bear
     \RR_{\mu\nu}  \eql -\alpha_0 \sigma_{,\mu}\sigma_{,\nu}, \label{Ei}\\
      \Box\sigma \eql 0                          \label{Esigma}
\ear
     where $\RR_{\mu\nu}$ and $\Box$ are the Ricci tensor and
     the D'Alem\-bert operator corresponding to $\gg_{\mu\nu}$.

     Vacuum $D$-dimensional equations are thus reduced to
     scalar-vacuum ones in 4 dimensions. Although such SAS configurations
     were repeatedly considered \cite{Rad77,RSh}, it makes sense to return
     to them to reveal some new features, in particular, those connected
     with higher dimensions.
\section{Field equations for axial symmetry}    
\markboth{Field equations for axial symmetry}
         {Field equations for axial symmetry}
        The SAS 4-metric in the Einstein gauge (\ref{GE}) may be
     written in the Weyl canonical form \cite{Synge}
\beq
        \ds= \e^{2\nu}dt^2-\e^{-2\nu}
                [\e^{2\beta}(d\rho^2 + dz^2) + \rho^2 d\phi^2]   \label{DsA} 
\eeq
        The field equations then can be written as
\bear
        \Delta\sigma \eql 0,                        \label{Esig}    \\   
        \Delta\nu    \eql 0,                        \label{Enu}     \\   
        \beta_z      \eql \rho(2\nu_\rho\nu_z
                       + \alpha_0\sigma_\rho\sigma_z)  \label{Bz}\\   
        \beta_\rho   \eql \rho[\nu_\rho^2
   - \nu_z^2 +\half\alpha_0\sigma_\rho^2 - \sigma_z^2)]\label{Brho}   
\ear
        where the indices $z$ and $\rho$ denote the partial derivatives
        $\partial_\rho$ and $\partial_z$, respectively, and $\Delta$ is the
        ``flat'' Laplace operator in the cylindrical coordinates:
\[
     \Delta =
     \rho^{-1}\partial_\rho(\rho\partial_\rho)+\partial_z\partial_z.
\]
        The integrability condition for (\ref{Bz}) and (\ref{Brho}) is
satisfied
        automatically.

        Following the example of \cite{Rad77}, let us seek solutions in the
        new coordinates $(x,y)$, connected with $\rho$ and $z$ by
\beq
     \rho^2 = L^2(x^2 +\varepsilon)(1-y^2),  \qquad
     z = Lxy                                  \label{XY}   
\eeq
     where $L$ is a fixed positive constant and $\varepsilon= 0,\ \pm 1$, so
     that $x$ and $y$ are spherical ($\varepsilon=0$), prolate
     spheroidal ($\varepsilon=-1$), or oblate spheroidal ($\varepsilon=+1$)
     coordinates, respectively. The Laplace operator $\Delta$
     acquires the form
\beq
        \Delta = \partial_x(x^2+\varepsilon)\partial_x
             +\partial_y(1-y^2)\partial_y.                \label{Delta} 
\eeq
     Separating the variables in Eq.(\ref{Enu}), i.e., putting $\nu
     (x,y) = \chi (x)\psi (y)$, one obtains
\bear
      {[(x^2 +
     \varepsilon)\chi_x]}_x + \lambda\chi \eql 0,  \label{Ex} \\ 
     {[(1-y^2)\psi_y]}_y - \lambda\psi     \eql 0   \label{Ey}    
\ear
        where $\lambda$ is the separation constant.
        Solutions to (\ref{Ey}) finite on the symmetry axis $\rho=0$
        are the Legendre polynomials $P_l (y)$, while
        $\lambda=l(l+1)$ with $l=0,1,2,\ldots$. The corresponding solutions to
        (\ref{Ex}) are combinations of Legendre functions of the first and
        second kinds.

        Eq.(\ref{Esig}) is solved in a similar way.

        This is the way to obtain solutions of arbitrary multipolarity $l$ or
        even superpositions of different multipolarities: after writing out
        the solutions to the linear equations (\ref{Enu}) and (\ref{Esig}),
        Eqs. (\ref{Bz}) and (\ref{Brho}) are integrable by
        quadratures. In what follows, however, we restrict ourselves to $l=0$
        (monopole solutions).
\section{Monopole solutions}     
\markboth{Monopole solutions}{Monopole solutions}
        The monopole solution to Eq.(\ref{Ey}) may without loss of
        generality be written in the form
\beq
     \e^{\psi} =[(1+y)/(1-y)]^{c_1},\quad c_1=\const.   \label{Psi}    
\eeq
        Regularity at $y = \pm 1$ then requires $c_1 =0$, so that $\nu=\nu(x)$.
        Eq.(\ref{Ex}) takes the form $(x^2+\varepsilon)d\chi/dx=\const$. Its
        integration leads to the following expressions for
        $\nu(x)$ satisfying the asymptotic flatness condition:
\beq
 \nu = \vars {-\half b \ln\frac{x+1}{x-1}, \quad &\varepsilon=-1,\\[2pt]
              -b/x,                              &\varepsilon=0,\\[2pt]
              -b \cot^{-1} x, \quad & \varepsilon =+1. } \label{Nu}   
\eeq
        In a similar way $\sigma(x)$ is found:
\beq
        \sigma = \vars {-\half s \ln\frac{x+1}{x-1}, \quad &\varepsilon=-1,\\
                           -s/x,                      &\varepsilon=0,\\
                     -s \cot^{-1} x, \quad & \varepsilon =+1. }
\label{Sigma}
\eeq
        Integrating (\ref{Bz}) and (\ref{Brho}), one obtains the
        expressions for $\beta(x,y)$ satisfying the asymptotic flatness
        condition $\beta(\infty,y)=0$
\beq
     \e^{2\beta} = \vars {
        (x^2-1)^K (x^2-y^2)^{-K},\ &\varepsilon=-1,\\
        \exp[-K(1-y^2)/x^2],   &\varepsilon=0,\\
        (x^2+y^2)^K(x^2+1)^{-K}, & \varepsilon=+1  }
                                                     \label{Beta}     
\eeq
        with $K = \half(2b^2 + \alpha_0 s^2) \geq 0$.
\section{General properties of the solutions}             
\markboth{General properties of the solutions}
         {General properties of the solutions}
        The solutions have been found under the boundary condition providing
        regularity (local euclidity) at the symmetry axis $\rho=0$, or $y=\pm
1$.

     At spatial infinity the solutions are
     asymptotically spherically symmetric. Indeed,
     assuming $y=\cos \theta$ where $\theta$ is the conventional polar
        angle, the SAS line element (\ref{DsA}) transformed by
        (\ref{XY}), is spherically symmetric under the condition
\beq
        \e^{2\beta} =   (x^2 + \varepsilon)/(x^2 + \varepsilon y^2).
                                                          \label{Sph} 
\eeq
        The condition (\ref{Sph}) holds for all the solutions in the limit
        $x\to\infty$ where they have Schwarzschild asymptotics.
        A particular expression for the Schwarzschild mass in terms of the
        integration constants is conformal gauge-dependent.
     Recalling that the mass is most meaningfully
     defined in the atomic gauge (\ref{GA}), one can write:
\bear
        \lal g^*_{tt}\approx 1 - 2GM/r, \qquad r\approx Lx;   \nn
     \lal GM = (b-s/4)L                            \label{Mass}     
\ear

        As for the whole space, the
        condition (\ref{Sph}) is fulfilled under the additional requirement
\beq
     K\varepsilon=\half(2b^2 + \alpha_0 s^2)\varepsilon =-1. \label{Sph1}
\eeq
        As $b$ and $s$ are real, this condition can hold only for
        $\varepsilon=-1$. Quite naturally, the solution with $\varepsilon=-1$
        constrained by (\ref{Sph1}) coincides with the well-known generalized
        Schwarzschild solution \cite{Sch}
        with the $(4+N)$-dimensional metric (\ref{DsD})
\bear
        ds_D^2 \eql {\kR}^{a_0} \nn
         \al{}- \al {\kR}^{-a_0-Na_1}\Bigl[dR^2+R^2\kR d\Omega^2\Bigr] \nn
         \lal \cm\cm {}+ {\kR}^{a_1} ds_1^2,           \nn
     \lal\nq Na_1^2 + a_0^2 + (a_0+Na_1)^2 =2              \label{Sph2}
\ear
        where the variable $R$ and the integration constants are connected with
        ours in the following way:
\beq
     x+1=R/k;\  Na_1=-s;\ a_0 =b+s/2;\ K=L.         \label{Sph3}
\eeq
     In [2] (see also references therein) solutions with a chain of
     Ricci-flat internal spaces, generalizing [11], are given; still
        more general spherical solutions with
        massless gauge and dilaton fields are discussed, e.g., in
     [9--13].

     The general solution with $\varepsilon=-1$ has a naked
     singularity at $x=1$ in all cases, except the
     spherically symmetric one when, in addition, the scalar field
        $\sigma$ is constant (or the extra dimensions are frozen), in agreement
     with [2]. The singularity at $x=1$ is anisotropic in all cases
        except (\ref{Sph2}): the metric coefficients behave in different ways
        when the singularity is approached from different directions. For some
        sets of integration constants the path to the singularity $y=\const,\
        \phi=\const,\ x\to 1$ has an infinite length; however, the explicit
        conditions of such a behavior are conformal gauge-dependent.

     In the case $\varepsilon=0$ the solution generalizes the well-known
     Curzon vacuum solution of general relativity \cite{Curz}, extensively
        studied in a number of papers, see, e.g., Ref.\cite{ScSz} and
        references therein. The metric can be written in the form (\ref{DsA})
        with
\beq
     \nhq \nu=-b/x, \ 2\beta =-K\rho^2/(L^2x^4),\ Lx=\sqrt{\rho^2+z^2}.
                                    \label{24}  
\eeq
        In the special case $s=0$ our solution coincides
        with the Curzon one up to a re-definition of constants.

        The solution is singular at $x=0$ in all cases except $s=b=0$ when it
        reduces to flat space-time. The singularity is anisotropic, such that
        even the finiteness or infiniteness of some metric coefficients can
        depend on the direction of approach. As shown in \cite{ScSz}, in the
        Curzon case the true nature of the singularity is revealed in some new
        coordinates, allowing one to penetrate beyond $x=0$ (in our notation).
        It turns out that curvature singularity $x=0$ has the shape of a ring
        and some spatial geodesics can pass through it to
        reach a second spatial infinity on the other side of the ring.

        This quasi-wormhole structure is preserved for the present,
        more general solution, although the exact conditions when this is the
        case or, on the contrary, $x=0$ is just a singular center, is
        conformal gauge-dependent.

        A further study of this solution, despite its possible interest, is
        beyond the scope of this paper. We will instead pay more attention to
        the solution with $\varepsilon=+1$, which has no curvature singularity
        and therefore seems more promising; and although a preferred conformal
        gauge does exist (the one in which the original $D$-dimensional
        theory is formulated), it is remarkable that the most important
        features of the configuration to be discussed do not depend on
        conformal factors of the form $\exp(\const\times\sigma)$.
\section{Membranes, strings and wormholes}  
\markboth{Membranes, strings and wormholes}
         {Membranes, strings and wormholes}
        The non-existence of a curvature singularity for
        $\varepsilon=+1$ does not necessarily mean
        that the space-time is globally regular. Let us study the limit
        $x\to 0$ in some detail.

        The functions $\nu,\ \sigma$ and $\e^\beta$ are finite at $x=0$.

        The curve $x=0,\ y=0$ as viewed in the coordinates $(\rho,z)$ lies in
     the plane $z=0$ and forms a ring $\rho=L$ of finite length (Fig.1). In
     the original conformal gauge (\ref{DsD}) the
     ring radius is $r_0 = L\exp(b\pi/2 + s\pi/4)$.

\unitlength=1mm
\begin{figure}
\begin{picture}(83,50)
        \put(0,0){\line(0,1){50}}
        \put(0,0){\line(1,0){83}}
        \put(83,0){\line(0,1){50}}
        \put(0,50){\line(1,0){83}}            
        \put(41.5,3){\vector(0,1){44}}
        \put(11.57,12){\circle*{2.5}}
        \put(71.5,12){\circle*{2.5}}
     \put(41.5,11){\circle*{1.25}}
     \put(41.5,13){\circle*{1.25}}
\thicklines
        \put(11.5,11){\line(1,0){60}}
        \put(11.5,13){\line(1,0){60}}
        \put(43,44){$z$}
        \put(9,6){$A$}
     \put(70,6){$B$}
     \put(6,16){\shortstack[l]{$x=0$\\ $y=0$\\ $\phi=\pi$}}
        \put(70,16){\shortstack[l]{$x=0$\\ $y=0$\\ $\phi=0$}}
        \put(42.5,16){\shortstack[l]{$x=0$\\ $y=+1$}}
        \put(42.5,4){\shortstack[l]{$x=0$\\ $y=-1$}}
\end{picture}
        \caption{\protect\small
     Axial section of the neighborhood of the ring $x=y=0$.
     The points $A$ and $B$, marked by big black circles,
     belong to the ring, the thick lines connecting them show the upper
     and lower disks $x=0,\ y{> \atop <}0$. }
\medskip \hrule
\end{figure}
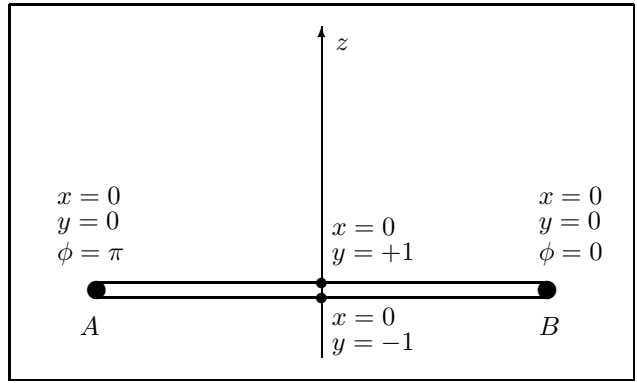

        The surface $x=0,\ y>0$ is a disk bounded by the
        above ring and parametrized by the coordinates $y$ and $\phi$.
        Its 2-dimensional metric is
\beq
        dl^2_{\rm disk}=L^2\e^{-2\nu-\sigma}
     [(1-\xi^2)^K d\xi^2+ \xi^2 d\phi^2]                 \label{Disk} 
\eeq
        where $\xi=\sqrt{1-y^2}$. This metric is flat if and only if $K=0$,
        i.e., when the solution is trivial. Otherwise the disk is curved
     but has a regular center at $y=1$ (the upper small black circle in
     Fig.1). The limit $x\to 0$ corresponds to approaching the disk from
     the half-space $z>0$.

        Another similar disk, the lower half-space one, corresponds to $y<0$.
        The two disks are naturally identified when our oblate
        spheroidal coordinates are used in flat space (obtained here in the
     case $K=0$).

        A possible identification of points $(x=0,\ y=y_0,\ \phi=\phi_0)$
     and $(x=0,\ y=-y_0,\ \phi=\phi_0)$, where $\phi_0$ is arbitrary and
        $0 < y_0 \leq 1$, leads to a finite discontinuity of the
        extrinsic curvature of the surfaces identified, or, physically, to a
        finite discontinuity of forces acting on test particles.
        Such a behavior corresponds to a membrane-like matter
        distribution. Thus a source of the global vacuum (or scalar-vacuum)
        gravitational field may be a membrane bounded by the ring $x=y=0$.

        There is another possibility, with
        no field discontinuity across the surface $x=0$. Namely, one
        can continue the $(x,y)$ coordinates to negative $x$ by just replacing
     in (\ref{Nu}) and (\ref{Sigma}) the function $\cot^{-1}x$ (undefined
     for $x<0$) by $\pi/2-\tan^{-1}x$, coinciding with the former at
     $x>0$.  This results in the appearance of another ``copy'' of the
     3-space, so that a particle crossing the regular disk $x=0$ along a
     trajectory with fixed $y$, threads a path through the ring and can
     ultimately get to another flat spatial infinity, with a different
     asymptotic value of $\nu$ and $\sigma$:
\bear
        \lal\nu(+\infty)=0,\quad \nu(-\infty)= -\pi b,\nn
     \lal\sigma(+\infty)=0, \quad \sigma(-\infty)= -\pi s,  \label{2As} 
\ear
        The function $\beta$ is even with respect to $x$ and hence coincides at
     both asymptotics. We obtain a wormhole configuration, nonsymmetric
     with respect to its ``neck'' $x=0$, having no curvature
     discontinuities, except maybe the ring $x=y=0$.

        It now remains to study the geometry near the ring. To this end let us
     consider a section of the ring by an $(x,y)$ surface at fixed $\phi$ and
        small $x$ and $y$. Its 2-dimensional metric near the point $x=y=0$ is
\beq
     dl^2_{(x,y)} = (x^2 + y^2)^{K+1}(dx^2 + dy^2).       \label{Ds2} 
\eeq
        This metric is flat, as is
        directly verified by the following transformation: introduce the polar
        coordinates $r$ and $\psi$ ($x=r\cos\psi$, $y=r\sin\psi$) and further
        transform them to $\xi$ and $\eta$ by the formulas
\beq
     r=[(K+2)\xi]^{1/(K+2)},\ \psi=\eta/(K+2).          \label{PolarT}
\eeq
        The result is
\beq
        dl^2_{(x,y)} = r^{2K+2}(dr^2 + r^2 d\psi^2) = d\xi^2 + \xi^2 d\eta^2
                                      \label{Polar} 
\eeq
     Thus we have above all assured that the metric near the ring
     $x{=}y{=}0$ is locally flat. However, it is locally flat on the ring
     itself only if the proper radius-circumference relation near the
     origin (the point $A$ or $B$ in Fig.1) in (\ref{Polar}) holds, i.e.,
     if $\eta$ is defined on a segment of length $2\pi$. Let us find out
     the $\eta$ range.

        Given $x>0$, the polar angle $\psi$ is defined on the segment
        $[-\pi/2,\ \pi/2]$, hence $\eta \in [-\pi -K\pi/2,\ \pi+K\pi/2]$.
        Consequently, the local flatness condition is fulfilled on the ring
     only in the trivial case $K=0$. Identifying
        the points  $\eta=\pm\pi$ and returning to
        the ($x,y$) coordinates, we then obtain flat space-time provided with
        oblate spheroidal coordinates with the single parameter $L$.

        For $x>0,\ K>0$ there is an excess polar angle, the situation
        opposite to a top-of-a-cone singularity. Such singularities are
        conventionally interpreted as cosmic strings, although in those objects
        a deficient rather than excessive polar angle range is considered.
        One can conclude that a possible source of the vacuum or scalar-vacuum
        gravitational field is a disk membrane bounded by a special kind of
        string.

        In the wormhole case $x$ can have either sign, hence
\beq
        \psi\in [-\pi,\ \pi]\
              \Rightarrow \eta\in [-(2+K)\pi,\ (2+K)\pi].  \label{Eta} 
\eeq
        Thus the axially symmetric wormhole solution contains a string-like
        ring singularity with a polar angle excess greater than $2\pi$.

        The excessive polar angle can have
        another mathematical meaning. Namely, if
        the excess is a multiple of $2\pi$, the singularity behaves like a
        branching point in a Riemannian surface of an analytic function of a
        properly defined complex variable. In our case the variable
        is $\zeta=\xi + i\eta$ and the analytic function is $\zeta^{1/(K+2)}$.
     Conformal mappings with analytic functions represent a natural way of
     regularizing metrics like (\ref{Ds2}); this method was indeed used in
     similar situations in [9, 10, 16, 17] where the relevant
     analytic function was logarithmic and the branching multiplicity was
     potentially (without additional identifications) infinite.

        If one postulates that the ``string'' should behave as a branching
        point, the integrality condition ($K={}$integer for
        (\ref{Eta})) is a quantization-type condition for the parameters of the
        solution. For instance, in the case $s=0$, i.e., a purely vacuum
        configuration (with maybe trivial extra dimensions), the mass is
     determined by $GM=bL$ and $K=b^2$, so that, given $L$ is a fixed
     length, the spectrum of masses has the form $GM=L\sqrt{K}$ where
        $K$ is a positive integer.

\section{Concluding remarks}                  
\markboth{Concluding remarks}{Concluding remarks}
        The results described appear from solving the field
        equations for pure vacuum or scalar vacuum in
        conventional general relativity as well as multidimensional
        gravity. One can conclude that SAS configurations can have
        nontrivial structures; those free of curvature
        singularities are in our view of greatest physical interest.
        Notably the singularities in SAS solutions are naked, except special
spherically
     symmetric cases (for the vacuum case see (\ref{Sph2}) for $a_1=0$).
     For general relativity this is a manifestation of the well-known
     uniqueness or no-hair theorems; it would be, however, of interest to
        analyze the situation in dilaton gravity for which spherically
        symmetric black-hole and non-black-hole solutions are known (see, e.g.,
     \cite{Br2}) and SAS ones are either known [10], or can be
        readily obtained, for instance, in $D$ dimensions with Ricci-flat
        internal spaces.

        The above solutions can be of interest for describing late
        stages of gravitational collapse and/or cosmological dark matter.
        Their monopole nature probably means that they cannot decay through
        gravitational-wave emission.

        Other generalizations of the present solutions, which are either known
        or easily obtainable by the known methods and are yet to be
        investigated in detail, are those with pure imaginary, nonminimally
        coupled and multiple scalar fields and/or multiple internal spaces.

\Acknow
     {This work was supported in part by the Russian Ministry of Science.}

\end{document}